\documentclass[aps,prl,twocolumn,groupedaddress,floatfix]{revtex4}
\usepackage{graphicx}
\usepackage{epstopdf}

\begin{document}

\title{Quasi-bound states in continuum}
\author{Hiroaki Nakamura}\email{hnakamura@nifs.ac.jp}
\affiliation{Department of Simulation Science, National Institute for Fusion Science,
Oroshi-cho, Toki, Gifu 509-5292, Japan}
\author{Naomichi Hatano}\email{hatano@iis.u-tokyo.ac.jp}
\affiliation{Institute of Industrial Science, University of Tokyo, Komaba, Meguro, Tokyo 153-8505, Japan}
\author{Sterling Garmon}
\author{Tomio Petrosky}
\affiliation{Center for Complex Quantum Systems, University of Texas at Austin, 1 University Station, C1609, Austin, TX 78712}

\date{\today}
\begin{abstract}
We report the prediction of quasi-bound states (resonant states with very long lifetimes) that occur in the eigenvalue continuum of propagating states for a wide region of parameter space.
These quasi-bound states are generated in a quantum wire with two channels and an adatom, when the energy bands of the two channels overlap.
A would-be bound state that lays just below the upper energy band is slightly destabilized by the lower energy band and thereby becomes a resonant state with a very long lifetime (a second QBIC lays above the lower energy band).
\end{abstract}



\maketitle

Since the bound state in continuum (BIC) was first proposed in 1929 by von Neumann and Wigner~\cite{vonNeumann29}, various researchers have reported its existence~\cite{Stillinger75,Gazdy77,Fonda60,Friedrich85,Capasso92,Deo94,Ordonez06,Bulgakov06}.
All studies agree that the phenomenon can only occur at discrete points of parameter space (i.e., the BIC is a zero-measure effect).

We here report the existence of a \textit{quasi}-bound state in continuum that exists over a wide region of parameter space (finite measure).
By quasi-bound state, we mean a resonant state with a very long lifetime, so long that it appears to be a bound state in space and hardly decays in time.
The quasi-bound state emerges when the system has an impurity level coupled to two overlapping energy bands with divergent van Hove singularities at the band edges.
If there were just one energy band, a bound state would appear just outside the edge (due to the singularity).
This would-be bound state is slightly destabilized by the other energy band, forming a quasi-bound state.
Since the quasi-bound state is virtually a bound state in continuum, it may be useful as a high-energy excited level of a carrier as a quantum device application.

\begin{figure}
\includegraphics[width=0.45\textwidth]{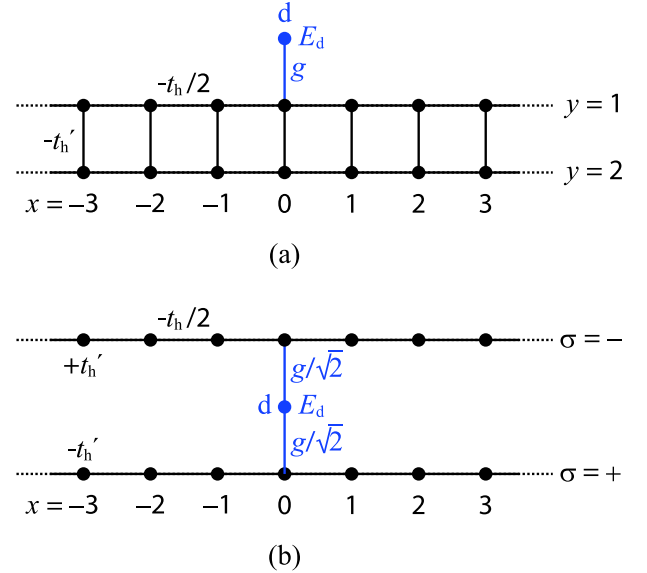}
\caption{(a) An adatom, or a dot, attached to a ladder. (b) After diagonalizing the ladder, the system is composed of a dot coupled to two independent channels.}
\label{fig0}
\end{figure}

The Hamiltonian for our system is a tight-binding model on a ladder with an adatom (Fig.~\ref{fig0}(a)):
\begin{eqnarray}\label{eq10}
\mathcal{H}&=&
-\frac{t_\mathrm{h}}{2}\sum_{y=1,2}\sum_{x=-\infty}^\infty
\left(c_{x+1,y}^\dag c_{x,y}+c_{x,y}^\dag c_{x+1,y}\right)
\nonumber\\
&&
-t'_\mathrm{h}\sum_{x=-\infty}^\infty
\left(c_{x,2}^\dag c_{x,1}+c_{x,1}^\dag c_{x,2}\right)
\nonumber\\
&&+g
\left(d^\dag c_{0,1}+c_{0,1}^\dag d\right)
+E_\mathrm{d}d^\dag d.
\end{eqnarray}
Here, $c_{x,y}^\dag$ and $c_{x,y}$ are the creation and annihilation operators of a spinless fermion at the site $(x,y)$ with integer $x$ ($-\infty<x<\infty$) and $y=1,2$, whereas $d^\dag$ and $d$ represent a dot, or an adatom, attached to the $(0,1)$ site of the ladder.
The first line of Eq.~(\ref{eq10}) gives the hopping matrix elements along the ladder, the second line the hopping elements across the ladder and the third line gives the hopping elements to and from the one-particle level of the dot.

The ladder has two eigenmodes $c_{x,\pm} \equiv (c_{x,1} \pm c_{x,2})/\sqrt{2}, $
which transform the Hamiltonian~(\ref{eq10}) to
\begin{eqnarray}\label{eq30}
\mathcal{H}&=&
-\frac{t_\mathrm{h}}{2}\sum_{\sigma=\pm}\sum_{x=-\infty}^\infty
\left(c_{x+1,\sigma}^\dag c_{x,\sigma}+c_{x,\sigma}^\dag c_{x+1,\sigma}\right)
\nonumber\\
&&
-t'_\mathrm{h}\sum_{\sigma=\pm}\sum_{x=\infty}^\infty
\sigma c_{x,\sigma}^\dag c_{x,\sigma}
\nonumber\\
&&+\frac{g}{\sqrt{2}}\sum_{\sigma=\pm}
\left(d^\dag c_{0,\sigma}+c_{0,\sigma}^\dag d\right)
+E_\mathrm{d} d^\dag d;
\end{eqnarray}
see Fig.~\ref{fig0}(b).
The system has two conduction channels
\begin{equation}\label{eq40}
c_{k_\pm,\pm}^\dag=\frac{1}{\sqrt{2\pi}}\sum_{x=-\infty}^\infty e^{ik_\pm x}c_{x,\pm}^\dag
\end{equation}
with $-\pi\leq k_{\pm} \leq \pi$, each of which forms an energy band
\begin{equation}\label{eq50}
\varepsilon_\pm(k_\pm)=-t_\mathrm{h}\cos k_\pm \mp t'_\mathrm{h}.
\end{equation}
Two divergent van Hove singularities occur in the density of states function at the edges of both bands \cite{06TGP}.
The two bands overlap (with one of the singularities of one band embedded within the continuum of the other) whenever $|t'_\mathrm{h}|<|t_\mathrm{h}|$.
The dot level couples to both channels as can be seen in Eq.~(\ref{eq30}).
Incidentally, we can also regard the Hamiltonian~(\ref{eq30}) as conducting electrons with spin $\sigma$ on a tight-binding chain under a magnetic field proportional to $t'_\mathrm{h}$.


In terms of the new operators for the wave number $k$ in Eq.~(\ref{eq40}), the Hamiltonian~(\ref{eq30}) takes the form of the coupled Friedrichs-Fano (Newns-Anderson) model Hamiltonian \cite{Frie, Sud, Fano,Ande}, which has been thoroughly studied.
Using the standard argument for this model, we obtain the dispersion equation for this coupled system:
\begin{eqnarray}\label{eq35}
&z - E_\mathrm{d} - \frac{g^2}{2}  
\left[
\frac{1}{ \sqrt{(z+ t'_\mathrm{h} )^2 - t_\mathrm{h}^2} }
-
\frac{1}{ \sqrt{(z- t'_\mathrm{h} )^2 - t_{\mathrm h}^2} }
\right]
= 0. & 	
\end{eqnarray}
This is equivalent to a twelfth order polynomial equation for $z$. 
The complex solutions of this equation correspond to the complex energy eigenvalues of the resonance states, with the decay rate given by the imaginary part.


Here, we focus on the numerical solution of this equation, using $t_{\rm h}^\prime = 0.345 t_{\rm h}$ and $g=0.1 t_{\rm h}$ for our demonstration. 
The twelve eigenvalues for $E_d = 0.3 t_{\rm h}$ are listed in Table I, as an example. 
\begin{table*}
\caption{The twelve discrete eigenvalues for $t'_\mathrm{h}=0.345t_\mathrm{h}$, $g=0.1t_\mathrm{h}$ and $E_\mathrm{d}=0.3t_\mathrm{h}$.}
\label{tab1}
\begin{tabular}{c||rl|rl|rl|c}
\hline\hline
state & $E/t_\mathrm{h}$ & & $K_{+}$ & &$K_{-}$ & & Riemann Sheet \\
\hline 
\textrm{P1(0.3)} &      1.34501152 &                                           &       3.14159265   & $+i$ 1.11593256                   &      3.14159265  & $+i$ 0.00480148  & I \\
\textrm{P2(0.3)}& $-$1.34500463 &                                          &                          & $+i$ 0.00304629                   &                        & $+i$ 1.11592751  & I \\
\hline
\textrm{Q1(0.3)} &      1.34501136 &                                           &       3.14159265   & $-i$ 1.11593245                   &      3.14159265  & $+i$ 0.00476787  & II \\
\textrm{Q2(0.3)} & $-$0.65501370 &  $-i$ 1.5093  $\times 10^{-7}$ &       1.25558888   & $-i$ 1.5875  $\times 10^{-7}$ & $-$0.00002882  & $+i$ 0.00523534  & II \\
\textrm{Q3(0.3)}  & $-$0.65501370 & $+i$ 1.5093  $\times 10^{-7}$ &  $-$1.25558888   & $-i$ 1.5875  $\times 10^{-7}$ &      0.00002882  & $+i$ 0.00523534  & II \\
\textrm{Q4(0.3)} &      0.29998854 &  $-i$ 0.00153774                   &       2.27180290   & $-i$ 0.00201224                   & $-$1.52576970  & $+i$ 0.00153930  & II \\
\textrm{Q5(0.3)} &      0.29998854 &  $+i$ 0.00153774                   &  $-$2.27180290   & $-i$ 0.00201224                   &      1.52576970  & $+i$ 0.00153930  & II \\
\hline
\textrm{R1(0.3)}  & $-$1.34500459 &                                           &                         & $+i$ 0.00303273                   &                        & $-i$ 1.11592748  & III \\
\textrm{R2(0.3)} &      0.65509906 & $-i$ 2.9331  $\times 10^{-6}$ &  $-$3.14138429   & $+i$ 0.01407702                   &      1.88609355  & $-i$ 3.0852  $\times 10^{-6}$ & III \\
\textrm{R3(0.3)}  &      0.65509906 & $+i$ 2.9331  $\times 10^{-6}$ &       3.14138429   & $+i$ 0.01407702                   & $-$1.88609355  & $-i$ 3.0852  $\times 10^{-6}$ & III \\
\hline
\textrm{S1(0.3)} &      0.29991927 &  $-i$ 0.01154476                   &       2.27161773   & $-i$ 0.01510419                   &      1.52570333  & $-i$ 0.01155625  & IV  \\
\textrm{S2(0.3)} &      0.29991927 &  $+i$ 0.01154476                   &  $-$2.27161773   & $-i$ 0.01510419                   & $-$1.52570333  & $-i$ 0.01155625  & IV 
\\
\hline\hline
\end{tabular}
\end{table*}
Each discrete eigenvalue can be distinguished by its position on the complex $K_+$ surface, the complex $K_-$ surface and the complex energy surface, where the three quantities are related by
\begin{equation}\label{eq60}
E=-t_\mathrm{h}\cos K_+ - t'_\mathrm{h}=-t_\mathrm{h}\cos K_- + t'_\mathrm{h}.
\end{equation}
The corresponding eigenfunction is given in the form
\begin{eqnarray}\label{eq70}
\lefteqn{
\left(\begin{array}{c}
\Psi(x,1;t) \\ \Psi(x,2;t)
\end{array}\right)
=}
\nonumber\\
&&e^{-iEt/\hbar} \left(
A_+e^{iK_+ |x|}
\left(\begin{array}{c}
1 \\ 1
\end{array}\right)
+A_-e^{iK_-|x|}
\left(\begin{array}{c}
1 \\ -1
\end{array}\right)
\right)
\qquad
\end{eqnarray}
with appropriate constants $A_\pm$.

The complex energy surface is composed of four Riemann sheets.
It is a physical requirement~\cite{Hatano07} that a resonant state decays in time but diverges in space.
The resonant state, therefore, has a negative imaginary part for its energy and a negative imaginary part for one or both of the wave numbers $K_+$ and $K_-$.
An eigenvalue in the upper $K_+$ plane and in the upper $K_-$ plane is defined to lay on Riemann sheet I and is a complete bound state; the states P1 and P2 in Table~\ref{tab1} are such states.
An eigenvalue in the lower $K_+$ plane and in the lower $K_-$ plane is defined to lay on Riemann sheet IV and is a complete resonant state; the state S1 in Table~\ref{tab1} is such a state.
(A state with a positive imaginary part of the energy is a so-called  anti-resonant state, which is the time reversal of a resonant state and hence grows in time; S2 is the time reversal state of S1.)

An eigenvalue in the lower $K_+$ plane but in the upper $K_-$ plane is defined to lay on Riemann sheet II, while one in the upper $K_+$ plane but in the lower $K_-$ plane is defined to lay on Riemann sheet III.
The resonant states on these two sheets (states Q2, Q4, and R2 in Table~\ref{tab1}) are the main focus of the present paper.

\begin{figure}
\includegraphics[width=0.33\textwidth,clip]{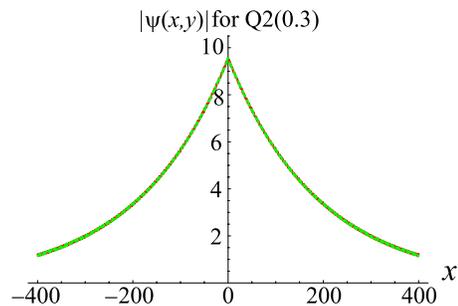}
\begin{center}(a)\end{center}
\includegraphics[width=0.45\textwidth,clip]{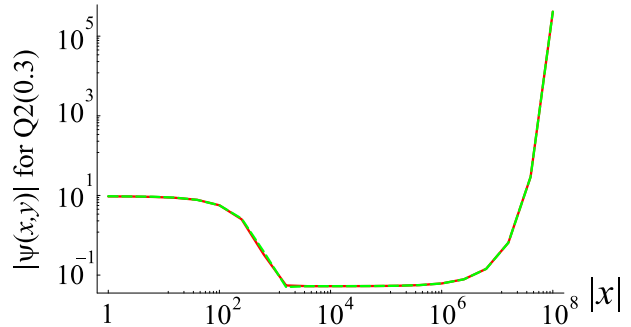}
\begin{center}(b)\end{center}
\caption{(a) The wave function modulus $|\psi(x,y)|$ of the state Q2 around the origin. (b) The same but away from the origin on the logarithmic scale.
The plots for $y=1$ (the upper leg) and $y=2$ (the lower leg) are almost indiscernible.
The parameters are set to $t'_\mathrm{h}=0.345t_\mathrm{h}$, $g=0.1t_\mathrm{h}$ and $E_\mathrm{d}=0.3t_\mathrm{h}$.}
\label{fig1}
\end{figure}

The resonant states in sheets II and III are resonant states because they diverge in space due to the negative imaginary part of only one of the wave numbers; see Eq. (\ref{eq70}).
Some of them, however, diverge in space very slowly and decay in time very slowly (the latter is demonstrated in Table~\ref{tab1}).
The state Q2, for example, \textit{appears} to be a localized state around the $x$-axis origin (Fig.~\ref{fig1}(a)) and exponentially diverges only far away (Fig.~\ref{fig1}(b)).
Indeed, solving the dispersion relation~(\ref{eq35}) by  perturbation expansion in $g$, we can show that the imaginary part of the energy of state Q2 is proportional to $g^6$ (extremely small for $g \ll 1$) due to the interaction between the divergent van Hove singularity in the upper energy band and the continuum of the lower band.
Since the real part of the energy lays within the lower energy band, we refer to this state as a quasi-bound state in continuum (QBIC).

In particular, the quasi-bound state Q2 has the real part of the energy just below the lower edge of the upper energy band $\varepsilon_-$, which occurs at $-t_\mathrm{h}+t'_\mathrm{h}=-0.655 t_\mathrm{h}$  in the present case.
If we had only the upper channel, this state would be a  bound state (due to the van Hove singularity) \cite{06TGP}.
In order to elucidate how the QBIC effect appears, we compare the above result to that obtained for the separate one-channel systems with the following new Hamiltonians with a structure similar to Eq.~(\ref{eq30}): 
\begin{eqnarray}\label{eq100}
\mathcal{H}_\pm&=&
-\frac{t_\mathrm{h}}{2}\sum_{x=-\infty}^\infty
\left(c_{x+1}^\dag c_{x}+c_{x}^\dag c_{x+1}\right)
\nonumber\\
&&+\frac{g}{\sqrt{2}}
\left(d^\dag c_{0}+c_{0}^\dag d\right)
+E_\mathrm{d}d^\dag d\mp t'_\mathrm{h}.
\end{eqnarray}
Here we set the coupling between the site $x=0$ and the adatom to $g/\sqrt{2}$ for quantitative comparison with the Hamiltonian~(\ref{eq30}).
We have also added the energy offset $\mp t'_\mathrm{h}$ so that the energy band of each Hamiltonian $\mathcal{H}_\pm$ may coincide with the energy bands $\varepsilon_\pm$ of the ladder system.
The Hamiltonian $\mathcal{H}_-$ thus mimics the upper energy band $\varepsilon_-$ of the $-$ channel of  the ladder system, while the Hamiltonian $\mathcal{H}_+$ mimics the lower band $\varepsilon_+$.

\begin{figure}
\includegraphics[width=0.3\textwidth]{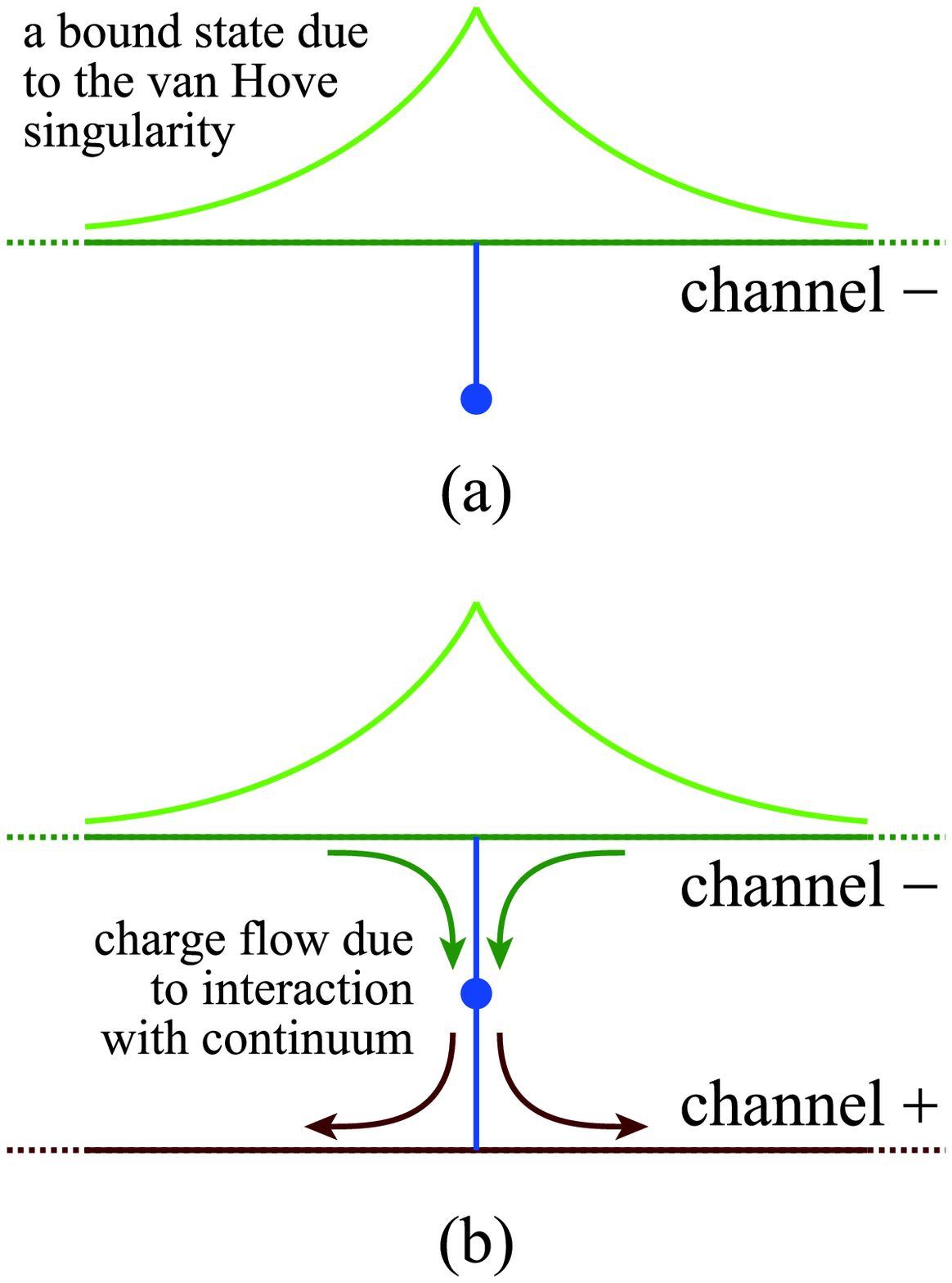}
\\
\vspace*{\baselineskip}
\includegraphics[width=0.38\textwidth]{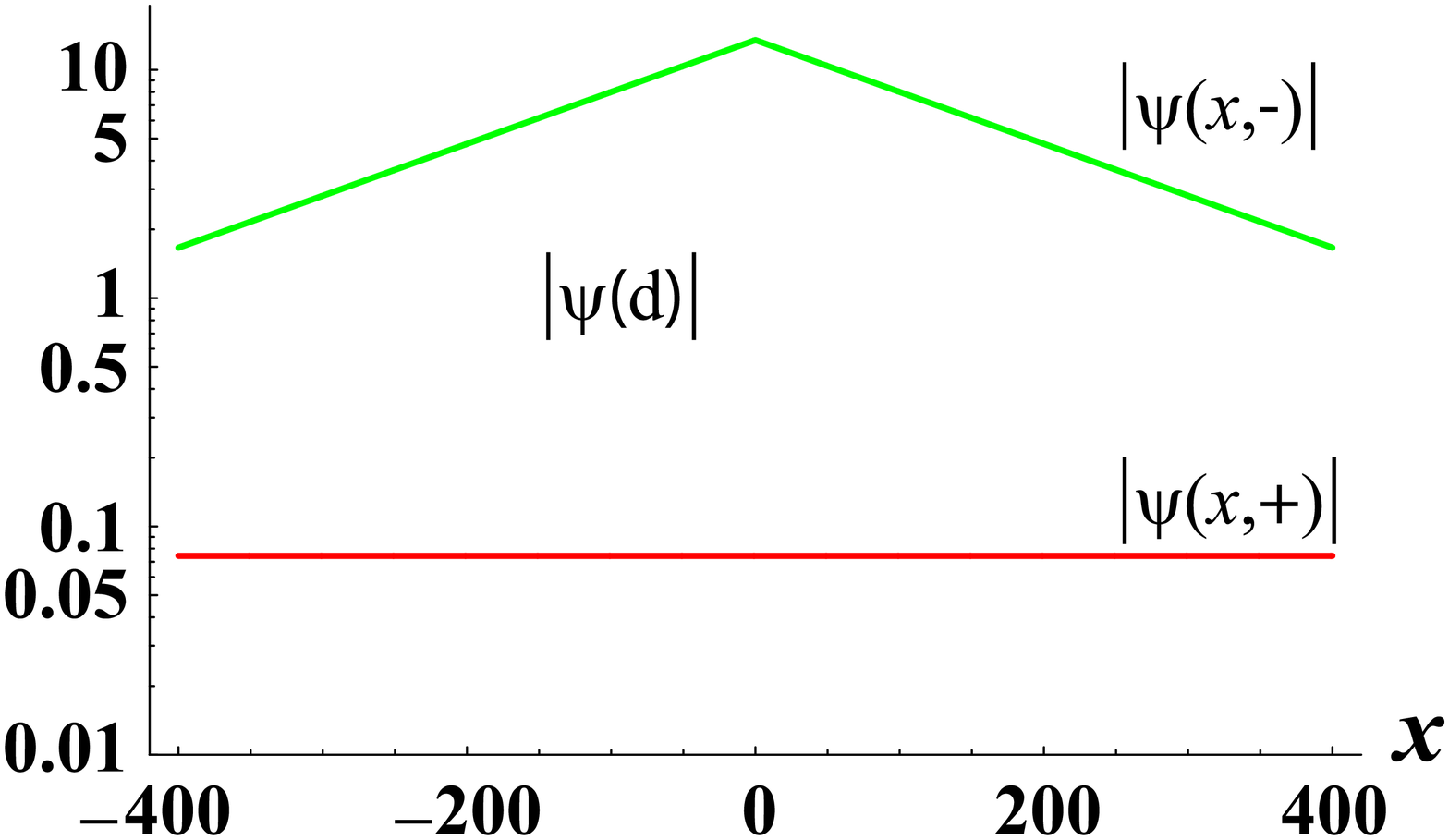}
\begin{center}(c)\end{center}
\caption{(a) A schematic view of the strongly bound state (due to the van Hove singularity) of a one-channel system with the eigenvalue just below the lower band edge.
(b) Some of the bound particles leak into the attached channel.
(c) The eigenfunction of the state Q2 for $t'_\mathrm{h}=0.345t_\mathrm{h}$, $g=0.1t_\mathrm{h}$ and $E_\mathrm{d}=0.3t_\mathrm{h}$.
The amplitude modulus of the $-$ channel, $|\psi(x,-)|$, that of the $+$ channel, $|\psi(x,+)|$, and that of the dot, $|\psi(\mathrm{d})|$, are indicated.}
\label{fig2}
\end{figure}

The bound state below the lower band edge of the Hamiltonian $\mathcal{H}_-$ for $t'_\mathrm{h}=0.345t_\mathrm{h}$, $g=0.1t_\mathrm{h}$ and $E_\mathrm{d}=0.3t_\mathrm{h}$ has a pure real energy $E/t_\mathrm{h} = -0.65501371$ and a pure imaginary wave number $K = i0.00523550$.
This bound state of the one-channel system (Fig.~\ref{fig2}(a)) indeed closely resembles the eigenvalue and the wave number $K_-$ of state Q2 of the two-channel system (see Table~\ref{tab1}).
When the second channel is attached to the one-channel system (Fig.~\ref{fig2}(b)), the energy is inside the conduction band of the $+$ channel so that a portion of the bound particles outside the edge of the $-$ channel leak into the $+$ channel and escape to infinity.
Notice $\mathop{\mathrm{Re}}K_-<0$, while $\mathop{\mathrm{Re}}K_+>0$; the particles on the channel $-$ are suck into the origin and spring out of the origin onto the $+$ channel.
This leak makes the state Q2 a resonant state;
it can be generally shown~\cite{Hatano07} that the imaginary part of the resonant eigenvalue arises due to the momentum flux escaping from the scattering potential.

\begin{figure}
\includegraphics[width=0.45\textwidth]{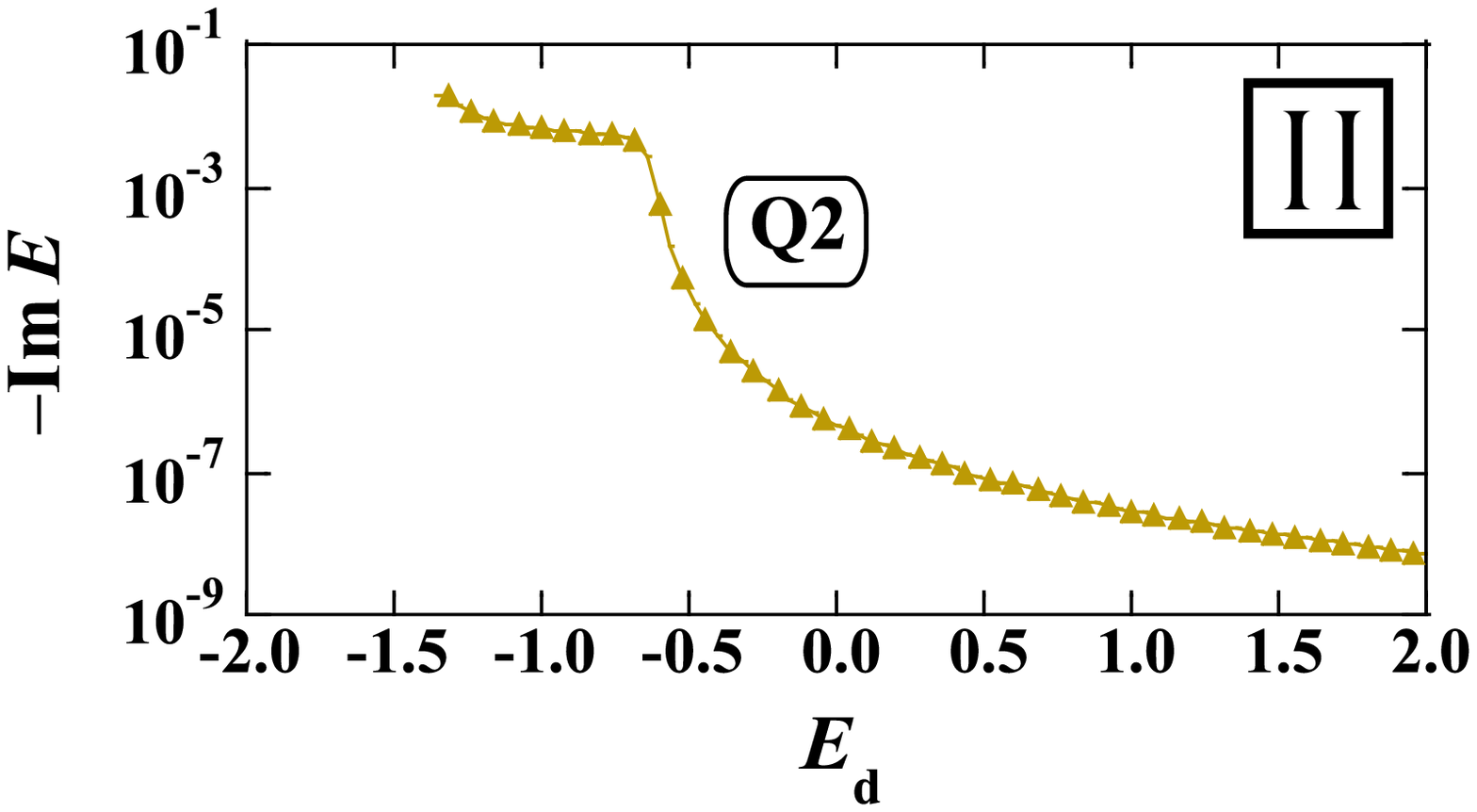}
\begin{center}(a)\end{center}
\includegraphics[width=0.45\textwidth]{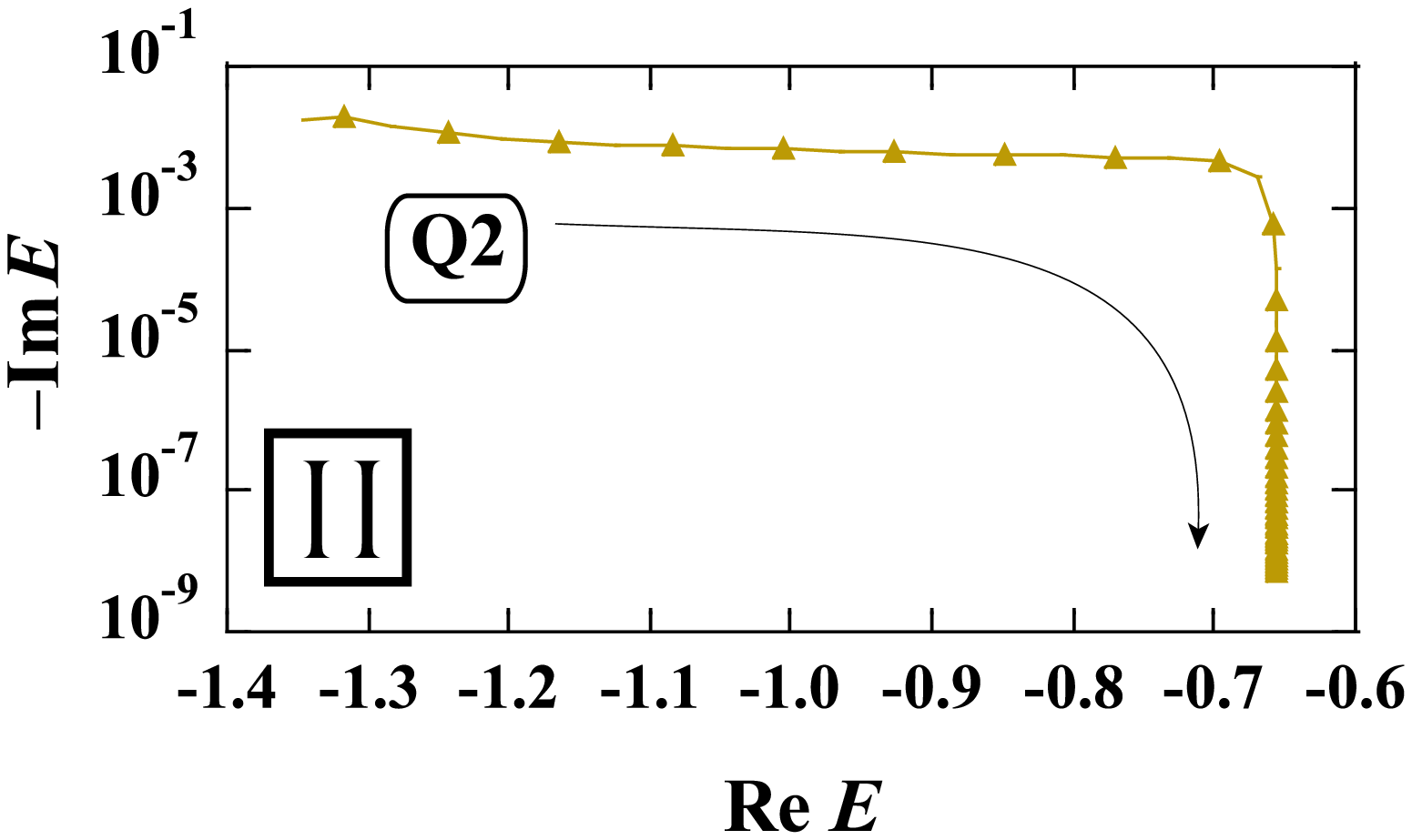}
\begin{center}(b)\end{center}
\caption{The imaginary part of the eigenvalues of the state Q2. (a) The dependence on the dot energy $E_\mathrm{d}$. The imaginary part vanishes for $E_\mathrm{d} \lesssim -1.345 t_\mathrm{h}$. (b) The dependence on the real part of the eigenvalue. The arrow indicates the increasing direction of $E_\mathrm{d}$. The range of $E_\mathrm{d}$ is the same as in plot (a).
The parameters are set to $t'_\mathrm{h}=0.345t_\mathrm{h}$ and $g=0.1t_\mathrm{h}$.}
\label{fig3}
\end{figure}

The actual eigenfunction of the state Q2 is exemplified in Fig.~\ref{fig2}(c).
One can show that the amplitude at the dot  $\psi(\mathrm{d})$ is order $g$ smaller ($g=0.1t_\mathrm{h}$ in this case) than the amplitude of $\psi(0,-)$ and the amplitude of $\psi(0,+)$ is order $g$ smaller than $\psi(\mathrm{d})$.
Hence the amount of the leak is small ($\sim g^2$).
The same analysis applies to the state R2, which is also a quasi-bound state.  Figure~\ref{fig3} shows that the quasi-bound states survive over a wide range of the parameter $E_\mathrm{d}$.

To summarize, we have found quasi-bound states in the continuum over a wide range of system parameters, which is in striking contrast to the bound state in the continuum.
The quasi-bound states emerge in the system with two overlapping eigenvalue continua when at least one of the bands contains a divergent Van Hove singularity.
For instance, a would-be bound state just below the upper continuum turns out to be a quasi-bound state in the lower continuum.
The simple picture of the mechanism suggests that the quasi-bound state may be found in various systems including a microwave tube with an ion~\cite{06TGP} and in the scattering of ions and nuclides.

\textit{Acknowledgments:}
The authors thank Prof. Satoshi Tanaka for useful discussions.
The work is supported partly by the National Institutes of Natural Sciences undertaking Forming Bases for Interdisciplinary and International Research through Cooperation Across Fields of Study and Collaborative Research Program (No.~NIFS06KDBT005, No.~NIFS06KDAT012, No.~NIFS07USNN002 and No.~NIFS07KEIN0091)
as well as by Grant-in-Aid for Scientific Research (No.~17340115 and No.~17540384) from the Ministry of Education, Culture, Sports, Science and Technology.
One of the authors (N.H) acknowledges  support by Core Research for Evolutional Science and Technology (CREST) of Japan Science and Technology Agency.

\end{document}